\documentclass[screen,nonacm,format=sigconf,balance=false]{acmart}

\usepackage[english]{babel}

\usepackage{amsmath,amssymb}

\usepackage{enumitem}

\theoremstyle{acmdefinition}

\theoremstyle{acmplain}

\usepackage{algorithmic}

\usepackage{float}
\usepackage{amsmath,stackengine, amssymb}

\usepackage{nicematrix}

\usepackage[ruled,vlined,linesnumbered]{algorithm2e}

\usepackage{tabularx}
\usepackage{tablefootnote}

\usepackage{mathtools}

\usepackage{tikz}
\usetikzlibrary{matrix,shapes,decorations.pathreplacing,backgrounds,positioning,calc,shapes.misc,mindmap}

\usepackage{ytableau}
\usepackage{booktabs}

\usepackage{subcaption}

\usepackage{graphicx}
\usepackage{multirow}
\usepackage{booktabs}
\usepackage{jlcode}

\usepackage{url}

\definecolor{myblued}{HTML}{0064FF}
\colorlet{softblue}{myblued!92!yellow}

\usepackage{color,svgcolor}
\usepackage{hyperref}
\hypersetup{%
        breaklinks=true,
}
\usepackage[hyphenbreaks]{breakurl}

\usepackage[nameinlink,capitalize,noabbrev]{cleveref}
\crefformat{footnote}{#2\footnotemark[#1]#3}
\Crefname{item}{Step}{Steps}
\crefname{item}{Step}{Steps}

\newcommand{\grobner}{Gr\"obner}
\newcommand{\faugere}{Faug\`ere}

\newcommand{\code}[1]{\texttt{#1}}
\newcommand{\codep}[1]{\texttt{#1}}

\renewcommand{\code}[1]{#1}

\newcommand{\best}[1]{\textcolor{softred}{#1}}

\newcommand{\smalltodo}[1]{}
\renewcommand{\smalltodo}[1]{{\color{red} TODO: {#1}}}

\definecolor{vlgrey}{HTML}{707070}
\newcommand{\zero}{\ensuremath{\textcolor{vlgrey}{0}}}

\newcolumntype{P}[1]{>{\centering\arraybackslash}p{#1}}

\definecolor{dblue}{HTML}{205488}
\definecolor{lblue}{HTML}{adc0d3}
\definecolor{dgreen}{HTML}{4F8355}
\definecolor{lgreen}{HTML}{BBCFBD}
\definecolor{yellow}{HTML}{F9EB0F}
\definecolor{red}{HTML}{FF4848}
\definecolor{greenalg}{HTML}{2cca2c}
\definecolor{ggrey}{HTML}{303030}
\definecolor{lgrey}{HTML}{404040}
\definecolor{vlgrey}{HTML}{707070}
\definecolor{nearwhite}{HTML}{DFDFDF}

\definecolor{mybl}{HTML}{62ABAB}

\definecolor{myblued}{HTML}{0064FF}

\definecolor{mybluedl}{HTML}{239BFF}
\definecolor{mybluel}{HTML}{697Fd4}
\definecolor{myredd}{HTML}{c31313}
\definecolor{myredl}{HTML}{c36767}
\definecolor{mygreend}{HTML}{2ca92c}
\definecolor{mygreenl}{HTML}{79a979}
\definecolor{myyellowd}{HTML}{F9EB0F}
\definecolor{mypinkd}{HTML}{E227A7}
\definecolor{myyellowl}{HTML}{e2a97c}
\definecolor{mybrown}{HTML}{442e0c}
\definecolor{mydpink}{HTML}{4d4d7c}
\definecolor{mylblue}{HTML}{338DF3}
\definecolor{mygreenyellow}{HTML}{ABF484}

\colorlet{softblue}{myblued!92!yellow}
\colorlet{softred}{myredd!90!yellow}
\colorlet{softgreen}{mygreend!90!yellow}

\begin{document}

\title{Groebner.jl:~A package for \protect \grobner{} bases computations in Julia}  
\author{Alexander Demin}
\affiliation{%
\institution{National Research University, \\
Higher School of Economics}
\country{Russia}}
\email{asdemin\_2@edu.hse.ru}

\author{Shashi Gowda}
\affiliation{%
\institution{Massachusetts Institute of Technology}
\country{United States}}
\email{gowda@mit.edu}

\begin{abstract}
We present \code{Groebner.jl}, a Julia package for computing Groebner bases with the F4 algorithm.
\code{Groebner.jl} is an efficient, portable, and open-source software.
\code{Groebner.jl} works over integers modulo a prime and over the rationals, supports basic multi-threading, and specializes in computation in the degree reverse lexicographical monomial ordering. The implementation incorporates various symbolic computation techniques and leverages the Julia type system and tooling, which allows Groebner.jl to compete with the existing state of the art, in many instances outperform it, and exceed them in extensibility. \code{Groebner.jl} is freely available at \url{https://github.com/sumiya11/Groebner.jl}.
\end{abstract}


\maketitle

\section{Introduction}
In 1965, Bruno Buchberger introduced \grobner{} bases, along with an elegant algorithm for their computation~\cite{classics}. Since then, \grobner{} bases have become a ubiquitous tool in computer algebra and acquired many scientific applications ranging from cryptography \cite{crypto} and analysis of dynamical systems and chemical reaction networks \cite{sianjulia,hamid} to robotics~\cite{robo}.
Algorithms that involve \grobner{} bases are implemented in major computer algebra systems, such as \linebreak \code{Maple} \cite{maple}, \code{Singular}~\cite{singular}, \code{Magma}~\cite{magma}, \code{CoCoA}~\cite{CoCoA,CoCoALib}, \code{Macaulay2}~\cite{M2}, and \code{Giac/Xcas}~\cite{giac}, to name a few.
Computing a \grobner{} basis, however, can be costly in time and space because of the natural complexity of the problem. Hence, it is often a bottleneck in applications.

A significant leap in performance was achieved with the introduction of the F4 algorithm by Jean-Charles \faugere{}~\cite{F4}. 
The F4 algorithm is a popular choice among practitioners, 
which can be explained by its intuitive design and high efficiency in general. Implementations of the F4 algorithm settled in the core of \grobner{} bases computation routines in many computer algebra systems. There also exist notable standalone implementations of F4, including the \code{FGb} project by \faugere{}~\cite{FGb}, and open-source alternatives such as \code{OpenF4}~\cite{openf4}, \code{gb}~\cite{gb}, and \code{msolve}~\cite{msolve}.

Julia~\cite{julia} is an actively developing interactive and expressive programming language with features well suited for scientific computing. Julia offers an array of packages for high-performance numerical and symbolic computation (e.g., SciML project~\cite{diffeq}, Oscar computer algebra system~\cite{OSCAR}). Nonetheless, there have been no competitive public implementations of \grobner{} bases algorithms written in Julia.

Julia community and research projects with implementations in Julia that depend on efficient \grobner{} bases computation, for example, \code{StructuralIdentifiability.jl}~\cite{SI}, mainly used the interface to \code{Singular} or the interface to the C library \code{gb} by Christian Eder as their back-end. These provide extensive functionality but come with several inconveniences, such as additional development and maintenance burden on the software authors for maintaining the aforementioned interfaces, limited flexibility
of possible user input, and, perhaps, most crucially, the lack of portability.

We engaged in the challenge of producing a native Julia package for competitive \grobner{} basis computations. As a result, we introduce \code{Groebner.jl}. \code{Groebner.jl} provides an efficient and tested \grobner{} bases computation routine based on the F4 algorithm. 
\code{Groebner.jl} can be used over integers modulo a prime and over the rationals, and specializes in computations in the degree reverse lexicographical ordering.
The package goes under the GNU General Public License, version 2, and works on all platforms that Julia supports (which include macOS, Windows, and Linux). \code{Groebner.jl} is compatible with the existing polynomial implementations, such as the ones from \code{AbstractAlgebra.jl}, \code{Nemo.jl}~\cite{Nemo.jl-2017}, and \code{DynamicPolynomials.jl} \cite{legat2021multivariatepolynomials}.

The heart of \code{Groebner.jl} is the implementation of the F4 algorithm over integers modulo a machine prime (i.e., a prime, which fits into a computer register). In F4, we use the normal selection strategy~\cite{F4}, and apply the Gebauer-M{\"o}ller criteria to detect redundant critical pairs~\cite{moller}. Parts of our implementation of F4 were adapted from the \code{msolve} library. More precisely, we adapt the routines for critical pair handling, symbolic preprocessing, and linear algebra. 
Our implementation over the rationals uses multi-modular computation and the two-stage learn and apply tracing strategy~\cite{tracingtrav}. On the apply stage, we perform linear algebra with a batch of prime numbers simultaneously, which provides further speed-ups.

The package documentation with the installation instructions, user interface description, and usage examples can be found on the project documentation page at \url{https://sumiya11.github.io/Groebner.jl}.
Additionally, we integrate our package into Symbolics.jl~\cite{symbolics} --- the primary Julia symbolic manipulation system. Groebner.jl can be used as a standalone Julia package or as an extension of Symbolics.jl.

The rest of the paper is organized as follows. In~\Cref{sec:algebraic}, we recall the basic algebraic background of \grobner{} bases. In~\Cref{sec:functionality}, we showcase the functionality of \code{Groebner.jl} using two simple examples. Some of our design decisions in the implementation are discussed in~\Cref{sec:implementation}. In~\Cref{sec:performance}, we compare the performance of \code{Groebner.jl} to computer algebra software \code{Singular} and \code{Maple}, a C++ library \code{OpenF4}~\cite{openf4}, and a C
library \code{msolve}~\cite{msolve}.
We conclude with a short discussion of the future work in~\Cref{sec:discussion}.

\section{Background}
\label{sec:algebraic}

We briefly recall basic notions on polynomial rings, \grobner{} bases, and the F4 algorithm. 
For an introduction to computational commutative algebra we refer to the classic works~\cite{cox,cca1-robbiano-kreuzer}.

\subsection{Notations and definitions}

Let $k$ be a field and let $\mathcal{R} \coloneqq k[x_1, \ldots, x_n]$ be the polynomial ring over $k$ in $n$ indeterminates. In this paper, $0 \in \mathbb{N}$. 
A monomial is a power-product of indeterminates $x_1^{a_1}\cdots x_n^{a_n}$ with $(a_1,\ldots,a_n) \in \mathbb{N}^n$, and a term is a monomial with a coefficient adjoined. The total degree of a monomial is the sum of its degrees.
Let $\mathcal{M}$ denote the set of monomials in $\mathcal{R}$.
A monomial ordering $<$ is a total ordering on $\mathcal{M}$, such that $u < v$ implies $w u < w v$ for all $u,v,w \in \mathcal{M}$. 
We consider only global monomial orderings (i.e., monomial orderings, for which $1 < x_i$ for all $1 \leqslant i \leqslant n$) ~\cite[Chapter 2, \S 2]{cox}. 

In this paper, we will mainly use the degree reverse lexicographical monomial ordering $<_{\operatorname{drl}}$. For monomials $u = x_1^{a_1} \cdots x_n^{a_n}$ and $v = x_1^{b_1} \cdots x_n^{b_n}$ with $a = (a_1,\ldots,a_n)$, $b = (b_1,\ldots,b_n)$,  $a,b \in \mathbb{N}^n$, we have $u <_{\operatorname{drl}} v$ if the total degree of $u$ is less than that of $v$, or they have the same total degree and the last non-zero entry of $a - b$ is positive~\cite[Chapter 1, \S 5, Definition 1.4.7]{cca1-robbiano-kreuzer}. For example, $x_1 x_3 <_{\operatorname{drl}} x_2^2$, as well as $x_3 <_{\operatorname{drl}} x_2 <_{\operatorname{drl}} x_1$.

With respect to a fixed monomial ordering, the largest monomial (term) of a polynomial $f \in \mathcal{R} \setminus \{0\}$ is called the leading monomial (term) of $f$, and we denote it by $\operatorname{lm}(f)$ (resp., $\operatorname{lt}(f)$). The leading monomial (term) is uniquely defined.
For a set of polynomials $F = \{ f_1, \ldots, f_m\} \subseteq \mathcal{R}$ the set $I = \langle F \rangle \subseteq \mathcal{R}$ defined as
\[
\langle F \rangle \coloneqq \{ h_1f_1 + \cdots + h_mf_m ~|~ ~h_i \in \mathcal{R}, 1 \leqslant i \leqslant m\}
\]
is called the ideal generated by $F$. The leading ideal of $I$ is said to be the set of terms $\operatorname{lt}(I) \coloneqq \{ \operatorname{lt}(f) ~|~ f \in I \}$. In general, $\operatorname{lt}(I)$ may be not equal to $\langle \operatorname{lt}(f_1), \ldots, \operatorname{lt}(f_m) \rangle$. 

\begin{definition}
A finite set of polynomials $G = \{g_1,\ldots,g_m \}$ is called a \grobner{} basis of an ideal $I$ with respect to monomial ordering $<$ if $G \subset I$ and $\operatorname{lt}(I) = \langle \operatorname{lt}(g_1), \ldots, \operatorname{lt}(g_m) \rangle$.
\end{definition}

Note that the definition of a \grobner{} basis depends on the choice of a monomial ordering. Every ideal in $\mathcal{R}$ has a \grobner{} basis~\cite[Chapter 2, \S 5]{cox}).

A reduction of a polynomial $f \in \mathcal{R}$ by a set of polynomials $G \subset \mathcal{R}$ is an operation of producing a new polynomial $f - c u g$,
where $c \in k \setminus \{0\}$, $u \in \mathcal{M}$, and $g \in G$ are such that $\operatorname{lt}(c u g)$ is contained in the set of terms of $f$. Such $c u g$ is called a reducer of $f$. We say that $f$ can be reduced to $f^*$ by $G$ if there exists a finite chain of reductions $f ~\rightarrow ~\cdots ~\rightarrow ~f^*$ by $G$.
For two polynomials $f, g \in \mathcal{R} \setminus \{0\}$, the S-polynomial of $f$ and $g$ is defined as
\[
\operatorname{Spoly}(f, g) \coloneqq \frac{\operatorname{lcm}(\operatorname{lm}(f), \operatorname{lm}(g))}{\operatorname{lt}(f)}f - \frac{\operatorname{lcm}(\operatorname{lm}(f), \operatorname{lm}(g))}{\operatorname{lt}(g)}g.
\] 
The pair of polynomials $f, g$ used in construction of an S-polynomial is often called a critical pair.

Buchberger gave an algorithmic definition of \grobner{} bases in terms of S-polynomials:
\begin{theorem}
    Let $I = \langle G \rangle$ be the ideal generated by a finite set of polynomials $G \subset \mathcal{R}$. Then, $G$ is a \grobner{} basis of $I$ with respect to monomial ordering $<$ if and only if the S-polynomial of $g_i, g_j$  for all $g_i, g_j \in G$ reduces to zero by $G$.
\end{theorem}

From a bird's-eye view, the algorithm proposed by Buchberger completes a generating set to a \grobner{} basis by iteratively constructing S-polynomials, reducing them with respect to the generating set, and appending any non-zero reductions back to the generating set, until no more non-zeros occur.

\begin{definition}
    Let $F = \{ f_1, \ldots, f_m \} \subset \mathcal{R}$. Let $T = \{t_1,\ldots,t_\ell\} \subset \mathcal{M}$ be the set of monomials that occur in $f_1,\ldots,f_m$, ordered from the largest to the smallest with respect to a fixed monomial ordering. The Macaulay matrix of $F$ is the matrix $M \in k^{k \times \ell}$ such that $M_{ij} = \operatorname{coeff}(s_i, t_j)$ for $1 \leqslant i \leqslant k, 1 \leqslant j \leqslant \ell$, where $\operatorname{coeff}(f, t)$ denotes the coefficient of $f$ in $t$.
\end{definition}
\begin{example}
    \label{ex:mac}
    Consider $F = \{ f_1, f_2 \}$ with $f_1 = x^2 + 2xz + 3z$ and $f_2 = 2x + 4y + 3$. We fix the monomial ordering $<_{\operatorname{drl}}$ and have $T = \{x^2, xz, x, y, z, 1\}$. The Macaulay matrix of $F$ is
    \[
    \begin{NiceTabular}{ccccccc}
       & $x^2$ & $xz$ & $x$ & $y$ & $z$ & $1$   \\
$f_1$  & $1$ & $2$ & $\zero$ & $\zero$ & $3$ & $\zero$\\
$f_2$  & $\zero$ & $\zero$ & $2$ & $4$ & $\zero$ & $3$
\CodeAfter \SubMatrix[{2-2}{3-7}]
\end{NiceTabular}.
    \]
\end{example}

\subsection{F4 algorithm}

As before, we fix a monomial ordering on $\mathcal{M}$, and let $F = \{f_1,\ldots,f_m\}$, $I = \langle F \rangle$. Consider the Macaulay matrix whose rows are the products $u f_i$ for all $u \in \mathcal{M}$ up to some fixed total degree, $1 \leqslant i \leqslant m$. If the products are taken to a degree that is large enough, then a \grobner{} basis can be extracted from the rows of this matrix in a row echelon form~\cite{matrixmacaulay,matrixmacaulay2}. This approach is mainly of theoretical interest, since the bound on the required degree is prohibitively large. Still, it may provide an insight into the workings of the F4 algorithm.

\begin{algorithm}
\caption{F4 algorithm (basic version)}
\label{alg:f4}
\begin{description}[itemsep=0pt]
\item[Input:] a set of polynomials $F = \{ f_1, \ldots, f_m\} \subset \mathcal{R}$
\item[Output:] a \grobner{} basis $G$ of $\langle F \rangle$
\end{description}

\begin{enumerate}[label = \textbf{(\roman*)}, 
leftmargin=*, align=left, labelsep=2pt, itemsep=4pt]
    \item Set $G \coloneqq F$ and let $P \coloneqq \{\operatorname{Spoly}(g_i, g_j) ~|~ g_i, g_j \in G\}$
    \item \label{step:alg1-3} While $P \neq \varnothing$ do
    \begin{enumerate}[label = (\alph*), ref = Step \theenumi (\alph*), leftmargin=*, align=left, labelsep=2pt, itemsep=4pt]
    \item\label{f4:select} Select $S \coloneqq \{s_1,\ldots,s_k\} \subseteq P$ and set $P \coloneqq P \setminus S$
    \item \label{f4:prep} Construct the Macaulay matrix $M$ from $s_1,\ldots,s_k$ and the reducers of $s_1,\ldots,s_k$ from $G$
    \item\label{f4:linalg} Compute $s_1^*,\ldots,s_k^*$, reductions of $s_1,\ldots, s_k$ \linebreak with respect to $G$, by echelonizing $M$
    \item For all $1 \leqslant i \leqslant k$, such that $s_i^* \neq 0$,\\
    set $G \coloneqq G \cup \{s_i^*\}$ and $P \coloneqq P \cup \{\operatorname{Spoly}(g, s_i^*)~|~g \in G\}$
    \end{enumerate}
    \item Return $G$
\end{enumerate}
\end{algorithm}

The F4 algorithm was introduced by \faugere{} to improve the computation of \grobner{} bases in practice~\cite{F4}. We outline its basic version in~\Cref{alg:f4}. 
Roughly speaking, instead of considering all polynomial multiples up to a large degree, the F4 algorithm uses S-polynomials to obtain a small Macaulay matrix, similarly to how the Euclidean algorithm does not require the whole Sylvester matrix to compute the gcd.

The process of constructing a Macaulay matrix in~\ref{f4:prep} is usually referred to as \textit{symbolic preprocessing}. 
The F4 algorithm amortizes the cost of symbolic preprocessing over a number of S-polynomials and replaces the costly multivariate polynomial arithmetic with linear algebra methods.

In practice, many of the S-polynomials in~\ref{f4:linalg} reduce to zero, which makes some rows in the matrix superfluous from the computational standpoint. Special rules were proposed to detect and discard such S-polynomials in advance. These include the two Buchberger's criteria~\cite{classics}, their development in~\cite{moller}, and more involved signature-based rules~\cite{F5,signaturesurvey}.
Furthermore, the order in which S-polynomials are selected in~\ref{f4:select} affects the performance in practice. There exist dedicated strategies that determine the preferred order of selection, perhaps the most common being the normal and sugar selection strategies~\cite{onesugarcube,learningstrat,slimgb}.

An echelon form of the Macaulay matrix in~\ref{f4:linalg} can be computed using straightforward Gaussian elimination. In practice, the Macaulay matrix is usually sparse, and it is customary to apply specialized combinations of dense and sparse methods~\cite{linalg1,perry}. When working over the rational numbers, the computation leads to heavy intermediate expression swell, and the use of multi-modular techniques becomes critical~\cite{slimgb,modularnold}.

The general idea of multi-modular algorithms is to solve the problem modulo several integers and reconstruct the answer, instead of solving directly over the rationals. In the F4 algorithm, this allows to control the size of entries in the Macaulay matrix.
When using the F4 algorithm in a multi-modular fashion, tracing turns out to be a relevant technique~\cite{tracingtrav}.
Tracing is a technique of accelerating the computation of a sequence of \grobner{} bases that come from specializations of the same ideal. Tracing consists of two main stages: during the \textit{learn} stage, in the first modular runs of the F4 algorithm, one learns which rows of the Macaulay matrices reduce to zero (and thus are redundant), and during the \textit{apply} stage, in the subsequent modular runs, one uses the knowledge to omit redundant rows from the matrices.

\section{Functionality}
\label{sec:functionality}

\code{Groebner.jl} implements the F4 algorithm and aims at providing an intuitive user interface for computing \grobner{} bases. Let us consider an example of a simple computation performed by our package.

To install \code{Groebner.jl}, type the following in a Julia session:

\begin{lstlisting}[columns=fullflexible]
import Pkg; Pkg.add("Groebner")
\end{lstlisting}

\code{Groebner.jl} does not provide an interface to manipulate polynomials.
Instead, the package should be used in combination with existing symbolic manipulation packages, for example, \code{AbstractAlgebra.jl} or \code{DynamicPolynomials.jl}, which can be installed analogously.
After the installation, the code to compute a \grobner{} basis over the rational numbers goes as follows:

\begin{lstlisting}[columns=fullflexible]
using Groebner, AbstractAlgebra  # load packages

R, (x, y, z) = QQ["x","y","z"]   # define a ring
F = [x^2*y + 2y^3, x*y^2 + 3z]   # define polynomials

G = groebner(F)                  # compute basis
\end{lstlisting}

The last command computes the reduced \grobner{} basis of the system over the rationals with respect the lexicographical monomial ordering with $x > y$. It will output something like the following:

\begin{lstlisting}
4-element Vector{AbstractAlgebra.Generic.MPoly{...}}:
 y^6 + 9//2*z^2
 x*z - 2//3*y^4
 x*y^2 + 3*z
 x^2*y + 2*y^3
\end{lstlisting}

\code{Groebner.jl} can be used over integers modulo a prime $p$\linebreak for $p < 2^{64}$ and over the rationals. 
Over the rationals, the default algorithm in \code{Groebner.jl} is Monte-Carlo probabilistic due to the use of multi-modular techniques.
\code{Groebner.jl} supports the lexicographical, the degree lexicographical, the degree reverse lexicographical, as well as weighted, block, and matrix monomial orderings.

\code{Groebner.jl} is already employed as the backend for \grobner{} basis computations in a number of Julia packages, including \code{StructuralIdentifiability.jl}~\cite{SI},
\code{SIAN-Julia}~\cite{sianjulia}, and in the implementation reported in~\cite{fibres24}.

\code{Groebner.jl} exports its internal multi-modular tracing to the user to unlock efficient implementation of algorithms that rely on multi-modular or evaluation-interpolation schemes~\cite{rur,interpolation,RODRIGUEZ2017342}. A common usage pattern consists in learning the trace of computation once, and applying it for computing similar \grobner{} bases. We showcase this functionality on the Katsura-10 system
: 

\begin{lstlisting}[columns=fullflexible]
using Groebner, AbstractAlgebra

# katsura over the integers in drl
kat = Groebner.katsuran(10, k=ZZ, ordering=:degrevlex)

# reduce the coefficients modulo two different primes
kat_zp1 = map(f -> change_base_ring(GF(2^30+3),f), kat)
kat_zp2 = map(f -> change_base_ring(GF(2^30+7),f), kat)

# learn a trace
trace, _ = groebner_learn(kat_zp1);

# using the trace, apply for another prime
success, gb = groebner_apply!(trace, kat_zp2);
\end{lstlisting}

This pattern becomes particularly appealing when the number of applications is large, since application is more efficient than an independent \grobner{} basis computation. 
To demonstrate this, we compare the runtime of the default F4 implementation in \code{Groebner.jl} against the runtime of application:

\begin{lstlisting}
using BenchmarkTools

@btime groebner($kat_zp2);
#  731.000 ms (84551 allocations: 139.38 MiB)

@btime groebner_apply!($trace, $kat_zp2);
#  263.576 ms (44002 allocations: 87.02 MiB)
\end{lstlisting}

For mode details, including an example of using the function \codep{groebner\_apply!} in batches to gain even more speed-up, we refer to the package documentation page at \url{https://sumiya11.github.io/Groebner.jl}.

\section{The implementation}
\label{sec:implementation}

\code{Groebner.jl} is written in Julia.
Julia uses multiple dispatch to support function overloading for different combinations of argument types with little to no running time overhead.
The design of Groebner.jl exploits this Julia feature.
In principle, our F4 implementation can be instantiated with any type of coefficient or monomial, as long as the types adhere to a common interface.
Hence, it is convenient to add a more specialized sub-program to the package to extend its capabilities.

Julia is a just-in-time (JIT) compiled language. Programs in Julia are compiled to LLVM IR, which is then compiled to native instructions by LLVM~\cite{LLVM:CGO04}. With some exceptions (which are discussed later in this section), we have observed that performance-critical code sections in Groebner.jl compile to compact and reasonable machine instructions across a variety of computer architectures. Thus, in linear algebra in \code{Groebner.jl}, we generally trust the compiler to apply the low-level optimizations such as SIMD and inlining.

The core of \code{Groebner.jl} is the implementation of the F4 algorithm. In F4, we use the normal selection strategy~\cite{F4}, and apply the Gebauer-M{\"o}ller criteria~\cite{moller} to detect redundant critical pairs.
Over the rationals, we use multi-modular techniques together with tracing.

\subsection{Polynomial representation}

Efficient polynomial representations for symbolic computation have been studied, e.g., by Monagan and Pearce~\cite{monoagan:pearce}. 
The polynomial representation in \code{Groebner.jl} is the distributed representation. A polynomial is represented with a pair of arrays: the array of monomials sorted with respect to the current monomial ordering and the array of coefficients. For example, the arrays $\{xy^2, x, z\}, \{1, 2, -1\}$ constitute the polynomial $xy^2 + 2x - z$ with respect to $<_{\operatorname{drl}}$. The arrays of coefficients are used interchangeably both as the coefficients of polynomials and as the rows of the Macaulay matrices in a sparse format. 

\code{Groebner.jl} implements two monomial representations: a classic exponent vector and a packed exponent vector. The packed representation is similar to the one from~\cite{packing}, where the monomial degrees are stored compactly within machine integers.
The packing strategy favors the degree reverse lexicographical ordering and allows monomial comparisons to be performed in a couple of word comparisons.
Our packed representation handles up to $31$ variables and total degrees up to $127$, which is sufficient for many examples.
We observed that the packed representation provides a total speed-up of 15\% for some problems and slightly reduces memory consumption.

We have experimented with sparse monomial representations and have found them not effective for problems with close to hundred variables due to costly sparse monomial arithmetic. When the number of variables is large, we use exponent vectors.

We have claimed above that Julia allows the package to be readily extended with a sub-routine to adapt it to a specific application, and the following example provides an illustration of this.
In \code{Groebner.jl}, the monomial representation is selected at the beginning of the computation. Then, the specialized functions that implement monomial arithmetic are just-in-time compiled and inlined by Julia.
In particular, this relocates the work from running time to compile time, a reasonable compromise considering the computational intensity of \grobner{} bases.

Monomials are kept in a hashtable that is instantiated for each \grobner{} basis computation. Whenever a new monomial is encountered during the computation, it is added to hashtable, and persists in hashtable throughout the entire computation. Thus, each monomial is stored in a single copy. 
One known trick is to use a hash function $h: \mathcal{M} \rightarrow \mathbb{N}$ such that $h(u) + h(v) = h(uv)$, for monomials $u,v \in \mathcal{M}$~\cite{parallel}. For example, $h$ can be a linear form on the exponent vectors. This way, when two monomials are multiplied, the hash of their product is obtained as the sum of their respective hashes.

\begin{figure}
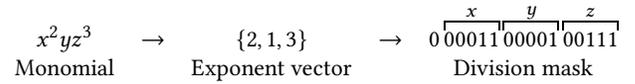

    \centering
\[ \begin{array}{cccccc}
\mbox{$x^2yz^3$} & \rightarrow & \mbox{$\{2, 1, 3\}$} & \rightarrow & \mbox{$0\overbracket[0.3pt]{00011}^x\overbracket[0.3pt]{00001}^y\overbracket[0.3pt]{00111}^z$} \\
\mbox{Monomial} & & \mbox{Exponent vector} &  & \mbox{Division mask}
\end{array}\] 
\caption{Monomial division mask. A monomial $x^2yz^3$ is represented by an array of integers $\{2, 1, 3\}$. The array, in turn, is compressed to a 16-bit machine integer, 5 bit per exponent.}
\label{fig:divmask}
\end{figure}

In symbolic preprocessing, monomial-by-monomial divisibility check is performed myriad times. We use division masks to accelerate divisibility checks, similar to \code{msolve}~\cite{msolve}. The idea is to keep an additional integer to store the monomial degrees compactly, as displayed in Figure~\ref{fig:divmask}. When the number of variables is large, we compress the division mask, so that a single bit in the mask may correspond to several variables.

If not overflowed, division mask allows checking monomial divisibility in a couple of word operations. Say, there are two monomials, $m_1, m_2 \in \mathcal{M}$, with division masks $d_1$ and $d_2$, respectively. Then, $m_2 | m_1$ is equivalent to $(\sim d_1)~\&~d_2 = 0$, where $\sim$ ~is the bitwise negation, and \& is the bitwise and.

\subsection{Linear algebra}
\label{secsec:linalg}

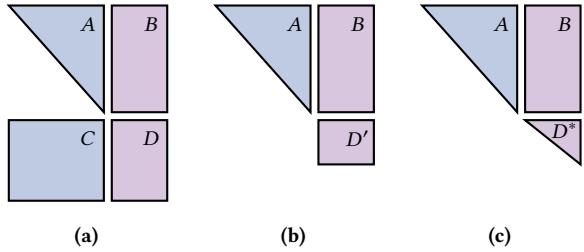
\begin{figure}
\centering
\begin{subfigure}[b]{0.15\textwidth}
\centering
\begin{tikzpicture}[
    >=latex,
    line width=1pt,
    tips=on proper draw
]
\matrix[
    matrix of nodes,
    text height=1.0ex,
    text depth=0.75ex,
    text width=1.29ex,
    align=center,
    column sep=4pt,
    row sep=4pt,
    nodes in empty cells,
] at (0,0) (M){ 
&   &   &   &  \\
&   &   &   &  \\
&   &   &   &  \\
&   &   &   &  \\
&   &   &   &  \\
};
\draw[thick,fill=ACMDarkBlue!20,draw] (M-1-1.center)
-- (M-3-3.south east)
-- (M-1-3.east)
-- cycle;
\draw[thick,fill=ACMDarkBlue!20,draw] (M-4-1.north)
-- (M-4-3.north east)
-- (M-5-3.south east)
-- (M-5-1.south)
-- cycle;
\draw[thick,fill=ACMPurple!20,draw](M-1-4.west)
-- (M-1-5.center)
-- (M-3-5.south)
-- (M-3-4.south west)
-- cycle;
\draw[thick,fill=ACMPurple!20,draw](M-4-4.north west)
-- (M-4-5.north)
-- (M-5-5.south)
-- (M-5-4.south west)
-- cycle;
\node at (M-1-3.south){\sf \small $A$};
\draw (M-1-5.south west) node {\sf \small $B$};
\draw (M-4-3.center) node {\sf \small $C$};
\draw (M-4-5.west) node {\sf \small $D$};
\end{tikzpicture}
\caption{}
\label{fig:matrix1}
\end{subfigure}
\begin{subfigure}[b]{0.15\textwidth}
\centering
\begin{tikzpicture}[
    >=latex,
    line width=1pt,
    tips=on proper draw
]
\matrix[
    matrix of nodes,
    text height=1.0ex,
    text depth=0.75ex,
    text width=1.29ex,
    align=center,
    column sep=4pt,
    row sep=4pt,
    nodes in empty cells,
] at (0,0) (M){ 
&   &   &   &  \\
&   &   &   &  \\
&   &   &   &  \\
&   &   &   &  \\
&   &   &   &  \\
};
\draw[thick,fill=ACMDarkBlue!20,draw] (M-1-1.center)
-- (M-3-3.south east)
-- (M-1-3.east)
-- cycle;
\draw[thick,fill=ACMPurple!20,draw](M-1-4.west)
-- (M-1-5.center)
-- (M-3-5.south)
-- (M-3-4.south west)
-- cycle;
\draw[thick,fill=ACMPurple!20,draw](M-4-4.north west)
-- (M-4-5.north)
-- (M-5-5.north)
-- (M-5-4.north west)
-- cycle;
\node at (M-1-3.south){\sf {\small $A$}};
\draw (M-1-5.south west) node {\sf {\small $B$}};
\draw (M-4-5.west) node {\sf {\small $D'$}};
\end{tikzpicture}
\caption{}
\label{fig:matrix2}
\end{subfigure}
\begin{subfigure}[b]{0.15\textwidth}
\centering
\begin{tikzpicture}[
    >=latex,
    line width=1pt,
    tips=on proper draw
]
\matrix[
    matrix of nodes,
    text height=1.0ex,
    text depth=0.75ex,
    text width=1.29ex,
    align=center,
    column sep=4pt,
    row sep=4pt,
    nodes in empty cells,
] at (0,0) (M){ 
&   &   &   &  \\
&   &   &   &  \\
&   &   &   &  \\
&   &   &   &  \\
&   &   &   &  \\
};
\draw[thick,fill=ACMDarkBlue!20,draw] (M-1-1.center)
-- (M-3-3.south east)
-- (M-1-3.east)
-- cycle;
\draw[thick,fill=ACMPurple!20,draw](M-1-4.west)
-- (M-1-5.center)
-- (M-3-5.south)
-- (M-3-4.south west)
-- cycle;
\draw[thick,fill=ACMPurple!20,draw](M-4-4.north west)
-- (M-4-5.north)
-- (M-5-5.north)
-- cycle;
\node at (M-1-3.south){\sf \small $A$};
\draw (M-1-5.south west) node {\sf \small $B$};
\draw (M-4-5.west)++(0,3pt) node {\sf \small $D^*$};
\end{tikzpicture}
\caption{}
\label{fig:matrix3}
\end{subfigure}
\caption{Macaulay matrix in \code{Groebner.jl}: (a) initially, after symbolic preprocessing (b) after reducing each row in the block $C|D$ with $A|B$
(c) after inter-reducing the rows of $D'$.}
\label{fig:matrices}
\end{figure}

The F4 algorithm performs polynomial reduction by transforming the Macaulay matrix into a row echelon form. The Macaulay matrices that appear in a typical run of F4 have special structure: they are sparse, partly triangular, and can be conveniently represented with a 4-block matrix, as shown schematically in~\Cref{fig:matrices}.

Initially, the matrix consists of two parts: the upper part with reducers (the block $A|B$ in~\cref{fig:matrix1}) and the lower part with S-polynomials (the block $C|D$). Symbolic preprocessing produces at most one reducer per column of the matrix, which means that the pivot rows in the block $A|B$ are readily located. Hence, we can arrange the rows of $A|B$ so that $A$ is upper-triangular.

We perform Gaussian elimination to obtain the reduction of $C|D$ with $A|B$, which gives us the matrix in~\cref{fig:matrix2}. Row reductions are performed using the sparsest rows first. Since $A$ is triangular, $C$ vanishes. At this point, the pivot rows in $D'$ are already known, and all other rows reduced to zero.

In the next step, the rows of $D'$ are inter-reduced to obtain the reduced row echelon form, as in the block $D^*$ in \cref{fig:matrix3}. The rows of $D^*$ correspond to reduced S-polynomials, which are potentially new elements of the basis.

Since many of the rows reduce to zero, in addition to the standard Gaussian elimination approach, we implement the probabilistic method attributed to Allan Steel from the Magma team (for details see~\cite{randomizedlinalg}).
Finally, we note that in \code{Groebner.jl} it is possible to print a minimized sparsity pattern of the Macaulay matrices by using the command \codep{groebner(system,loglevel=-3)}.

We implement the arithmetic in $\mathbb{Z}/p\mathbb{Z}$ in three different flavors. All three store the elements of $\mathbb{Z}/p\mathbb{Z}$ as machine integers, and the bit size of integers varies depending on the size of $p$.
The first flavor, the \textit{generic} arithmetic, takes remainder modulo a prime after each multiplication in $\mathbb{Z}/p\mathbb{Z}$. To speed-up division by $p$, it precomputes the numbers $m, s \in \mathbb{Z}$ such that $\left \lfloor{\frac{x}{p}}\right \rfloor = \left \lfloor{\frac{m\cdot x}{2^s}}\right \rfloor$ for all $x$ representable by a machine integer~\cite[Chapter 10]{hacker}. This effectively replaces division with a multiplication and a shift. For finding suitable $m$ and $s$ we adapt the implementation from the Julia standard library\footnote{This improvement was proposed by Chris Elrod in private communication.}.

The second version of arithmetic in $\mathbb{Z}/p\mathbb{Z}$ uses the \textit{signed} strategy discussed in~\cite{parallel}.
This implementation stores the coefficients as signed machine integers. To subtract the product of machine integers $c, y$ from $x$, it writes (using a ternary operator)
\begin{align*}
    x &\coloneqq x - cy,\\
    x &\coloneqq x < 0 ~? ~x + p^2 : x,
\end{align*}
where the addition of $p^2$ ensures that $x$ does not overflow. In our implementation, the generic arithmetic and the signed arithmetic attain similar performance.

The third implemented strategy is the \textit{delayed} modular arithmetic. The idea is to exploit the spare leading zero bits in the machine representation of $p$ and delay the modulo operation to just until the overflow is about to happen. 
In the most common case of primes that fit into 32-bit  words, we set the cut-off point at 27 bits: if the prime fits in 27 bits, we opt to use delayed arithmetic; otherwise, we use the generic arithmetic. 

\begin{table}
\centering
\caption{Breakdown of runtime over integers modulo $2^{30}+3$.}
    \begin{tabular}{l|rrrrr}
    & Cyclic-9 & Eco-14 & Goodwin (w.) & Yang1\\
    \hline
    Pair selection & $4.07$ & $1.95$ & $20.57$ & $1.68$\\
    Symbolic prep. & $8.34$ & $8.64$ & $5.90$ & $5.62$\\
    Linear algebra & $242.03$ & $168.83$ & $284.01$ & $7.73$ \\
    Pair update & $6.71$ & $2.03$ & $79.50$ & $10.80$\\
    Auto-reduction & $4.67$ & $44.6$ & $0.0$ & $0.19$\\
    Other & $0.47$ & $1.40$ & $0.09$ & $0.98$\\
    \end{tabular}
  \label{table:distr}
\end{table}

In~\cref{table:distr}, we present a breakdown of the time spent in different parts of our F4 implementation over integers modulo $2^{30}+3$ in the degree reverse lexicographical monomial ordering. All timings are in seconds. For the description of our benchmarking environment, we refer to~\cref{sec:performance}.
For each example system, we report the time spent in: selecting critical pairs, symbolic preprocessing, echelonizing the matrix, updating the set of critical pairs, and auto-reducing the final basis. The table captures $>99\%$ of the total runtime for each of the examples.

In three out of the four examples, the runtime is dominated by linear algebra. In the Goodwin (w.) system, the largest Macaulay matrix is of size $403,677 \times 374,837$ with $41,698,725$ non-zero elements ($0.02756\%$). We note that while symbolic preprocessing corresponds to a smaller fraction of total runtime, it causes the most memory allocations.

\subsection{Multi-modular computation}

Over the rational numbers we apply modular computation techniques. The algorithm selects multiple machine primes and computes a \grobner{} basis over integers modulo a prime for each of them. It then reconstructs the results to the rationals using the Chinese reminder theorem and rational number reconstruction~\cite[Chapter 5, \S10]{tolstaya}. Finally, the correctness of the reconstruction is assessed. 

In the multi-modular runs, we use 31-bit primes, store the coefficients in 32-bit machine integers, and carry out the arithmetic on 64-bit machine integers.
We compute the modular images in batches, and the size of the batch does not exceed 15\% of the total number of already computed images. We attempt to reconstruct a small subset of the coefficients, and, if successful, proceed with the reconstruction of the rest\footnote{This improvement was proposed by Fabrice Rouillier in private communication.}. Our rational number reconstruction routine uses Nemo.jl~\cite{Nemo.jl-2017}, which calls Flint~\cite{flint}. We also experimented with replacing rational reconstruction with the Maximal Quotient Rational Reconstruction~\cite{MQRS}, and it was not effective on our benchmarks. 

The multi-modular approach is Monte-Carlo probabilistic: there is no out of the box guarantee that the reconstructed basis is correct over the rational numbers. One needs to apply a costly correctness check to assure that~\cite{modularnold}. We do not perform a full check,  but use a relaxed randomized check and a heuristic. 
The randomized check verifies that the recovered basis over the rationals is a \grobner{} basis modulo a randomly chosen prime number. The heuristic checks that the sizes of coefficients in the recovered basis are not too large compared to the square root of the size of the modulo.

\subsection{Learn and apply in batches}
\label{sec:trace}
Our multi-modular computation uses tracing~\cite{tracingtrav,msolve,giac-tracing}. 
The classic tracing consists of two stages: the learn stage and the apply stage. The learn stage records useful meta-data in the first modular runs of the F4 algorithm, and the apply stage reuses this data in the subsequent runs. On the apply stage, our implementation mirrors the F4 computation, but does not instantiate the rows of the matrix that were deemed redundant. It also caches some auxiliary data across the modular runs, such as the order of rows and columns in the matrices.

The apply stage in \code{Groebner.jl} is capable of operating in batches. The idea is to share the linear algebra step in the F4 algorithm between a set of primes. Let $p_1,\ldots,p_N$ be some fixed machine primes. Instead of performing the computation separately for each of them, we ``glue'' the polynomial coefficients together and compute over the $N$-tuples $\mathbb{Z}/p_1\mathbb{Z} \times \cdots \times \mathbb{Z}/p_N\mathbb{Z}$.
This amortizes the cost of other parts of the F4 algorithm, and opens opportunities for SIMD parallelization. 
Since our data structures and the main F4 routine are generic with respect to the type of coefficient, this functionality was straightforward to implement.

\begin{table}
\centering
\caption{Runtime of apply stage in batches with 31-bit primes.}
\resizebox{0.48\textwidth}{!}{
    \begin{tabular}{l|rrrrr}
     & Classic F4 & Learn & Apply & Apply (4x) & Apply (8x)\\
    \hline
    Cyclic-8 & $1.20$ & $3.75$ & $0.53$ & $0.76$ & $4.48$\\
    Cyclic-9 & $98.96$ & $691.77$ & $75.67$ & $141.23$ & $277.85$\\
    Eco-13 & $9.26$ & $98.25$ & $4.07$ & $6.49$ & $15.39$\\
    Eco-14 & $97.60$ & $926.51$ & $55.31$ & $104.95$ & $287.67$\\
    Katsura-11 & $5.66$ & $49.95$ & $2.19$ & $3.84$ & $11.35$\\
    Katsura-12 & $45.52$ & $523.29$ & $15.96$ & $28.73$ & $62.91$\\
    Cholera & $107.16$ & $376.82$ & $51.06$ & $117.03$ & $224.11$\\
    SEIRP & $237.51$ & $1307.65$ & $125.60$ & $308.66$ & $597.90$
    \end{tabular}}
  \label{table:apply}
\end{table}

In~\cref{table:apply}, we report on the running time of the classic F4, the learn and apply stages (the columns Learn and Apply), and the apply stage done in batches of sizes $4$ and $8$ (the columns Apply (4x) and Apply (8x)). All timings are in seconds. 

The first thing to note is that the non-batched apply is in all instances faster than the classic F4. Furthermore, the ratio of the runtime of application in batches of size $4$ to the runtime of non-batched application is $1.43 < \frac{\text{Apply (4x)}}{{\text{Apply}}} < 2.44$. This suggests that batched application can save up to an amortized factor of $2.7$ on some of these examples. With the batch size $8$, the benefit is not as apparent, and we note that we did not made attempts at optimizing this code yet.

Our modular computation uses the batched apply scheme with the size of the batch $N=4$. The arithmetic on the tuples $\mathbb{Z}/p_1\mathbb{Z} \times \mathbb{Z}/p_2\mathbb{Z} \times 
\mathbb{Z}/p_3\mathbb{Z} \times
\mathbb{Z}/p_4\mathbb{Z}$ delegates arithmetic operations to each element of the tuple, which then use one of the strategies discussed in~\Cref{secsec:linalg}. In this case, the Julia compiler autonomously applies SIMD optimization on all Intel CPUs that we tried, but was observed to struggle on ARM architectures (e.g., on Apple M2 Max). 

We also experimented with larger batch sizes, such as $N=16$ or $N=32$, and have found no significant speed-ups over $N=4$. It is possible that one would need to adjust the overall linear algebra scheme in F4 to benefit from larger batch sizes.

In theory, the use of the batched strategy in multi-modular computation could increase the peak memory usage in case one of the intermediate polynomials in the Macaulay matrix is very long. We have not observed such situations in practice, which can be perhaps explained by the fact that on the apply stage the superfluous rows in the matrix are pruned.

\subsection{The use of multi-threading}

Some of the functions in \code{Groebner.jl} use multi-threading.
For computations over integers modulo a prime, we parallelize linear algebra routines by performing Gaussian elimination in parallel. 
Over the rationals, we parallelize the multi-modular runs, and do not parallelize linear algebra.

\begin{table}
\centering
\caption{Runtime over the rationals with multiple threads.}
    \begin{tabular}{l|rrrr}
    Threads & $1$ & $4$ & $8$ & $16$\\
    \hline
    HIV2 & $17.61$ & $8.74$ & $7.07$ & $7.55$\\
    Chandra-12 & $95.65$ & $83.90$ & $87.70$ & $90.01$\\
    Reimer-8 & $550.30$ & $240.51$ & $220.95$ & $221.49$\\
    \end{tabular}
  \label{table:par}
\end{table}

We benchmark the speed-up obtained from parallelization over the rationals and report the results in~\Cref{table:par}.
All timings are real times in seconds.
While using multiple threads results in a noticeable improvement, the gain is far from optimal. The following factors could be in play: (1) the learn stage of the multi-modular computation and the reconstruction to the rationals are not parallelized; on the Chandra-12 example, reconstruction takes around $2/3$ of the time (2) the multi-modular computation makes many memory allocations, which stresses the Julia's Garbage Collector (GC).

We use the high-level macro \codep{Base.Threads.@threads} from the Julia standard library and the \code{Atomix.jl} Julia package for parallelizing our code.
For benchmarking, we start Julia with the command line option \codep{-\--threads=N} and compute the bases using the command \codep{groebner(system, threaded=:yes)} in \code{Groebner.jl}.

\section{Benchmarks}
\label{sec:performance}

In this section we will demonstrate the performance of our implementation on a set of benchmark problems\footnote{The sources of benchmark problems are available at \url{https://github.com/sumiya11/Groebner.jl/blob/54c9d6435438f1c8113d9393cace469e5b9b11bf/src/utils/examples.jl} and \url{https://github.com/sumiya11/Groebner.jl/tree/54c9d6435438f1c8113d9393cace469e5b9b11bf/benchmark/generate/benchmark_systems}}.
Benchmark suite consists of various real-world polynomial systems that differ in their specific hardness, like reduction process, critical pair handling, etc. Among polynomial systems chosen for comparison we have:
\begin{itemize}
    \item System solving examples: Chandra-n, Cyclic-n, Eco-n, \linebreak Henrion-n, Katsura-n, Noon-n, Reimer-n, Hexapod, IPP. 
    \item \code{SIAN-Julia} examples~\cite{sianjulia}: Cholera, Goodwin (w.), HIV2, NF-$\kappa$B (w.), SEIRP. The polynomials in these systems originate from differential ideals and contain a sizeable number of variables.
    \item \code{StructuralIdentifiability.jl} examples~\cite{SI}: SIWR, SEAIJRC. The equations of these systems take several MB on the disk, and the resulting bases are very simple.
    \item BioModels repository examples~\cite{biomodels}: BIOMD103, BIOMD123. These represent the steady-state ideals of systems of ordinary differential equations obtained from chemical reaction networks.  
    \item Some other examples~\cite{EDER20211}: bayes148, gametwo2, jason210, mayr42, yang1, alea6.
\end{itemize}

\begin{table}
\centering
\caption{Comparison of Groebner.jl with other software over integers modulo $2^{30}+3$ in $<_{\operatorname{drl}}$, all timings in seconds. The best running time in each row is highlighted in \best{red}.}
    \resizebox{0.48\textwidth}{!}{%
    \begin{tabular}{c|rrrrr}
    System &
    Groebner.jl &
    Maple &
    msolve &
    OpenF4 &
    Singular\\
    \hline
    Cyclic-8 & $1.46$ & \best{$1.23$} & $1.44$ & $9.43$ & $-$\\
    Cyclic-9 &  \best{$259.19$} & $340.73$ & $270.84$ & $5,551.77$ & $-$\\
    Eco-13 &  \best{$10.98$} & $13.18$ & $19.39$ & $74.54$ & $-$\\
    Eco-14 & \best{$213.55$} & $216.66$ & $259.84$ & $1,106.28$ & $-$\\
    Eco-15 & $2,115.58$ & \best{$1,851.99$} & $2,407.44$ & $-$ & $-$\\
    Henrion-7 & \best{$2.83$} & $3.15$ & $3.55$ & $29.81$ & $94.75$\\
    Henrion-8 & $871.57$ & $1,190.94$ & \best{$563.32$} & $-$ & $-$\\
    Katsura-11 &  $9.20$ & \best{$7.72$} & $8.90$ & $61.22$ & $1,387.56$\\
    Katsura-12 & $80.44$ & \best{$52.66$} & $76.77$ & $297.78$ & $-$\\
    Katsura-13 & $381.16$ & $691.95$ & \best{$267.54$} & $2,859.98$ & $-$\\
    Noon-8 &  $2.00$ & $1.95$ & \best{$1.87$} & $18.20$ & $4.93$\\
    Noon-9 &  $19.49$ & \best{$18.67$} & $22.32$ & $198.10$ & $47.89$\\
    Noon-10 &  \best{$155.23$} & $171.69$ & $178.75$ & OOM & $604.41$ \\
    Reimer-8 & \best{$31.75$} & $31.81$ & $41.38$ & $252.20$ & $-$\\
    Reimer-9 & \best{$2,043.98$} & $4,198.73$ & $4,022.78$ & $-$ & $-$\\
    Cholera & \best{$70.65$} & $98.30$ & $176.09$ & $-$ & $-$\\
    Goodwin (w.) & \best{$298.46$} & $368.19$ & $522.61$ & $-$ & $-$\\
    HIV2 & \best{$3.66$} & $4.05$ & $15.92$ & $202.62$ & $-$\\
    NF-$\kappa$B (w.) & \best{$325.64$} & $515.50$ & N/A\tablefootnote{Segmentation fault} & $-$ & $-$\\
    bayes148 &  $79.99$ & $54.41$ & $75.96$ & OOM & \best{$7.48$}\\
    gametwo2 & $23.84$ & $32.76$ & \best{$22.08$} & $38.92$ & $-$\\
    jason210 & $6.83$ & $11.68$ & $10.23$ & OOM & \best{$2.39$}\\
    mayr42 & $122.10$ & \best{$94.48$} & $118.97$ & N/A\tablefootnote{Crashed with ``terminate called after throwing an instance of 'std::length\_error'''} & $314.92$\\
    yang1 & \best{$27.20$} & $54.17$ & $71.36$ & OOM & $101.44$
    \end{tabular}}
  \label{table:first-ff}
\end{table}

We compare the single-threaded performance of \code{Groebner.jl}, version 0.7.0, with the \grobner{} basis computation routines in other software, which are also run in single-thread:
\begin{itemize}
    \item \code{Maple}~\cite{maple}: a general purpose computer algebra system. We use Maple 2021 and the command \codep{Basis} with the option \codep{fgb}~\cite{FGb}.
    \item \code{msolve}~\cite{msolve}: an open-source implementation of F4 in C. We build the library from source with default options\footnote{Commit hash fdfeaac7d3c128ba9840a4d212bcaeb12b237ce2.}. Over the rationals, we use the command \codep{msolve -g 2 -c 0}. Over integers modulo a prime, we also add the option \linebreak \codep{-l 44}, since we found it to be faster. We output the resulting basis to \codep{/dev/null} to minimize the influence of the I/O.
    \item \code{OpenF4}~\cite{openf4}: an open-source implementation of F4 in C++. We build the library from source with default options\footnote{Commit hash 7155c6ec4e5f776e0e12202716dc7da58469e77b.} and use the command \codep{f4}.
    \item \code{Singular}~\cite{singular}: an open-source computer algebra system. We use the command \codep{std} in Singular version 4.3.3, which uses the Buchberger's algorithm, accessed via the Julia package Singular.jl.
\end{itemize}

For each software in the list, we validate the correctness of a computed \grobner{} basis by comparing it to the output of the rest of the software. Storing a \grobner{} basis for future checks could take a lot of disk space, so we do not do that. Instead, we obtain a short SHA-256 certificate of each computed \grobner{} basis and store/compare the certificates.

All the timings in this section are elapsed times measured on Xeon(R) Gold 6130 CPU @ 2.10GHz 64-Core processor with AVX512 and 200 GB of RAM running Linux 11. 
Our benchmarking script is available at \url{https://github.com/sumiya11/Groebner.jl/tree/master/benchmark}.
For Julia, we do not take into account the JIT compilation time.

In~\Cref{table:first-ff}, we present the running times in seconds of \grobner{} basis computation over integers modulo $2^{30}+3$ in the degree reverse lexicographical ordering. In \code{Groebner.jl}, we use the command \codep{groebner} with default options. We report the runtime of Groebner.jl, and the runtimes of \code{Maple}, \code{msolve}, \code{OpenF4}, and \code{Singular}. In the table, we write ``OOM'' if the computation used more than 100 GB of RAM, and ``$-$'' if the computation was stopped after waiting roughly 10 times the runtime of \code{Groebner.jl}.

We observe that \code{Groebner.jl} outperforms \code{OpenF4} in all instances.
The ratio of the runtime of \code{Groebner.jl} to the runtime of \code{msolve} is within $[0.51, 1.42]$, which is a reasonable range. Groebner.jl seems to be slightly more optimized for \code{SIAN-Julia} systems with many variables. We note that the Buchberger's algorithm in \code{Singular} \linebreak demonstrates good performance on some classic examples.

\begin{table*}
\caption{Comparison of Groebner.jl with other software over the rationals in $<_{\operatorname{drl}}$. The best overall time is highlighted in \best{red}.}
\centering
    \resizebox{0.9\textwidth}{!}{%
    \begin{tabular}{cr|rrrr|rrrrr}
 \multirow{2}{*}{System} & \multirow{2}{*}{\# primes} & \multicolumn{4}{c|}{Groebner.jl, single modular run} & \multicolumn{4}{c}{Overall} \\ & & Classic F4 & Learn & Apply & Apply (4x) &
    Groebner.jl &
    Maple &
    msolve &
    Singular \\
    \hline
    Chandra-10 & $98$ & $0.13$ & $0.23$ & $0.02$ & $0.07$ & \best{$3.98$} & $25.70$ & $4.55$ & $44.94$ \\
    Chandra-11 & $110$ & $0.48$ & $1.15$ & $0.05$ & $0.12$ & \best{$15.26$} & $129.58$ & $21.31$ & $341.61$ \\
    Chandra-12 & $150$ & $2.68$ & $7.36$ & $0.45$ & $1.11$ & $\best{95.65}$ & $1,052.14$ & $104.74$ & $1,688.00$ \\
    Chandra-13 & $166$ & $15.06$ & $49.39$ & $0.84$ & $2.16$ &  $\best{527.64}$ & $4,409.00$ & $554.96$ & $-$ \\
    Cyclic-7 & $26$ & $0.08$ & $0.18$ & $0.03$ & $0.04$ & \best{$0.95$} & $1.41$ & $1.13$ & $-$ \\
    Cyclic-8 & $54$ & $1.17$ & $3.64$ & $0.50$ & $0.72$ & \best{$19.67$} & $23.82$ & $26.10$ & $-$ \\
    Eco-11 & $14$ & $0.32$ & $1.12$ & $0.14$ & $0.21$ & \best{$2.69$} & $4.87$ & $3.72$ & $-$ \\
    Eco-12 & $18$ & $2.08$ & $9.40$ & $0.85$ & $1.28$ & \best{$24.53$} & $35.06$ & $29.19$ & $-$ \\
    Eco-13 & $26$ & $8.84$ & $88.53$ & $3.58$ & $5.68$ & $288.08$ & $496.48$ & \best{$269.46$} & $-$ \\
    Henrion-7 & $342$ & $1.93$ & $5.29$ & $0.71$ & $1.13$ & \best{$347.42$} & $2,214.00$ & $391.25$ & $-$ \\
    Katsura-9 & $34$ & $0.19$ & $0.63$ & $0.06$ & $0.11$ & $3.17$ & $9.00$ & \best{$2.27$} & $201.00$ \\
    Katsura-10 & $54$ & $0.76$ & $4.92$ & $0.30$ & $0.51$ & $18.65$ & $84.81$ & \best{$17.54$} & $2,864.00$ \\
    Katsura-11 & $78$ & $4.99$ & $46.82$ & $1.99$ & $3.44$ & $187.59$ & $1318.31$ & \best{$167.57$} & $-$ \\
    Noon-8 & $6$ & $1.34$ & $1.56$ & $0.19$ & $0.33$ & \best{$4.90$} & $9.23$ & $11.37$ & $12.21$ \\
    Noon-9 & $6$ & $12.49$ & $13.06$ & $1.32$ & $2.29$ & \best{$30.11$} & $98.78$ & $91.26$ & $184.11$ \\
    Reimer-7 & $30$ & $0.77$ & $1.32$ & $0.10$ & $0.16$ & $9.08$ & $32.33$ & \best{$6.68$} & $-$ \\
    Reimer-8 & $78$ & $17.86$ & $42.05$ & $1.70$ & $2.69$ & $550.30$ & $2,472.00$ & \best{$256.74$} & $-$ \\
    Alea6 & $418$ & $0.14$ & $0.25$ & $0.06$ & $0.08$ & $43.63$ & $229.41$ & \best{$28.16$} & $-$ \\
    Hexapod & $1,102$ & $0.00$ & $0.01$ & $0.00$ & $0.00$ & $4.88$ & $63.68$ & \best{$3.42$} & $300.39$ \\
    IPP & $2,166$ & $0.01$ & $0.01$ & $0.00$ & $0.00$ & $47.85$ & $290.35$ & \best{$13.22$} & $858.20$ \\
    SIWR & $26$ & $0.39$ & $0.55$ & $0.01$ & $0.03$ & $1.64$ & $6.18$ & $2.08$ & \best{$0.63$}\\
    SEAIJRC & $34$ & $2.51$ & $3.07$ & $0.03$ & $0.10$ & $8.34$ &  $85.26$ & $13.55$ & \best{$6.33$}\\ 
    BIOMD103 & $342$ & $0.21$ & $0.60$ & $0.07$ & $0.12$ & $24.97$ & $168.87$ & \best{$22.80$} & $-$ \\
    BIOMD123 & $826$ & $0.80$ & $0.96$ & $0.16$ & $0.17$ & \best{$140.91$} & $3,291.37$ & $364.63$ & $-$ \\
    Cholera & $98$ & $50.50$ & $146.34$ & $20.13$ & $35.09$ & $1,245.15$ & \best{$527.64$} & $1,841.00$ & $-$\\
    Goodwin (w.) & $150$ & $430.81$ & $309.54$ & $34.96$ & $114.39$ & $2,282.27$ & \best{$2,120.47$} & $2,240.00$ & $-$ \\
    HIV-2 & $166$ & $2.60$ & $2.98$ & $0.24$ & $0.33$ & \best{$14.58$} & $23.85$ & $67.79$ & $-$\\
    SEIRP & $58$ & $76.31$ & $409.59$ & $41.65$ & $80.22$ & $1,682.00$ & $331.20$ & $2,451.00$ & $-$\\
    \end{tabular}}
  \label{table:rationals}
\end{table*}

Our next benchmark compares the software over the rational numbers. For each benchmark model,~\Cref{table:rationals} contains:

\begin{itemize}
    \item The number of 31-bit primes used in the multi-modular computation by \code{Groebner.jl}.
    \item The \code{Groebner.jl}-specific information about a single modular run. We give the timings in seconds for the independent computation (Classic F4), the learning of the trace (Learn), applying using the known trace (Apply), and applying on a batch of 4 primes at once (Apply (4x)). 
    \item The overall runtime of \code{Groebner.jl}, \code{Maple}, \code{msolve}, and \code{Singular}. \code{OpenF4} is not included for the comparison since it does not support computation over the rationals.
\end{itemize}

From the table, we can conclude that Groebner.jl is more than competitive over the rationals in single thread, in some examples beating Maple by an order of magnitude.
However, both Groebner.jl and msolve perform worse than Maple for some SIAN-Julia examples, and both lose to Singular on some systems from StructuralIdentifiability.jl. Notably, there exist examples where Groebner.jl is faster than msolve by a factor of three, and vice versa.

The information on a single modular of Groebner.jl in~\Cref{table:rationals} supports our findings in~\Cref{sec:trace}. It shows that the ratio of runtimes (Apply (4x)) / (Apply) is less than $4$ in all instances, and often is much closer to $2$.

We note that \code{Groebner.jl}, \code{Maple}, and \code{msolve} do not certify the \grobner{} basis over the rationals by default~\cite{maple}, and we do not know if \code{Singular} does.
Finally, we note that the some discrepancies in the running times between~\Cref{table:apply},~\Cref{table:first-ff}, and~\Cref{table:rationals} are caused by fluctuations in running times and by the fact that in~\Cref{table:rationals}, in a single modular run of Groebner.jl, we report the minimum of times across several runs to make the timings more stable.

\section{Concluding remarks}
\label{sec:discussion}

We presented a package for computing \grobner{} bases in Julia. The package works over integers modulo a prime and over the rationals, and specializes in computation in the degree reverse lexicographical monomial ordering. We observed that \code{Groebner.jl} is competitive with state of the art software, and in some instances outperforms it.
Some work could be necessary to measure and decrease the memory requirements of Groebner.jl.

In addition, we suggested a method of boosting the performance of multi-modular tracing by computing bases in small batches. It is possible that increasing the batch size would be beneficial. Indeed, in linear algebra in F4, a larger batch size would amortize the costs of dealing with sparsity in the matrix. The bulk of the work would be focused on performing useful computation on dense tuples of coefficients with predictable memory access pattern. A natural extension of this could be the use of GPUs to perform these computations, and we may explore this direction in the future.

\subsection*{Acknowledgements}
AD would like to thank Gleb Pogudin, Thomas Sturm, \linebreak Hamid Rahkooy, Michael Stillman, Fabrice Rouillier, Joris van der Hoeven, Chris Rackauckas, Vladimir Kuznetsov, and \linebreak Oleg Mkrtchyan for helpful discussions.

AD is grateful to The Max Planck Institute for Informatics and The MAX team at The Computer Science Laboratory of {\'E}cole polytechnique for providing computation resources.

\bibliographystyle{elsarticle-num} 
\bibliography{main}

\end{document}